\documentclass[prb,reprint]{revtex4-1}
\usepackage{graphicx}% Include figure files
\usepackage{bm}% bold math
\usepackage[english]{babel}
\usepackage{amsmath}
\usepackage{SIunits}
\usepackage{epstopdf}

\begin{document}
%___________________________________________________________________________________________
%
% title
%___________________________________________________________________________________________
\title{Role of MgO barriers for spin and charge transport in Co/MgO/graphene non-local spin-valve devices}
%___________________________________________________________________________________________
%
% authors
%___________________________________________________________________________________________
\author{F. Volmer}
\author{M. Dr\"{o}geler}
\author{E. Maynicke}
\author{N. von den Driesch}
\author{M. L. Boschen}
\author{G. G\"{u}ntherodt}
\author{B. Beschoten}
\thanks{e-mail: bernd.beschoten@physik.rwth-aachen.de}
\affiliation{II. Institute of Physics, RWTH Aachen University, 52056
Aachen, Germany} \affiliation{JARA-Fundamentals of Future
Information Technology, J\"{u}lich-Aachen Research Alliance,
Germany}

\date{\today}
%___________________________________________________________________________________________
%
% abstract
%___________________________________________________________________________________________
\begin{abstract}

We investigate spin and charge transport in both single and bilayer graphene non-local spin-valve devices. An inverse dependence of the spin lifetime $\tau_s$ on the carrier mobility $\mu$ is observed in devices with large contact resistance area products ($R_cA>\unit{1}{\kilo\ohm\micro\meter\squared}$). Furthermore, we observe an increase of $\tau_s$ with increasing $R_cA$ values demonstrating that spin transport is limited by spin dephasing underneath the electrodes. In charge transport, we measure a second contact-induced Dirac peak at negative gate voltages in devices with larger $R_cA$ values demonstrating different transport properties in contact covered and bare graphene parts. We argue that the existence of the second Dirac peak complicates the analysis of the carrier mobilities and the spin scattering mechanisms.

\end{abstract}
\maketitle

Graphene has drawn strong attention because of measured spin diffusion length of some $\mu$m at room temperature. While most spin transport devices only exhibit spin lifetimes up to several hundred picoseconds at room temperature~\cite{Tombros2007,Dlubak2012,PhysRevLett.107.047207,
doi:10.1021/nl301567n,doi:10.1021/nl301050a,PhysRevB.87.075455,PhysRevB.87.081405,PhysRevB.84.075453,
Neumann2013} there are only few reports with spin lifetimes above one nanosecond.\cite{PhysRevB.87.081402,doi:10.1021/nl2042497,PhysRevLett.107.047206} Nevertheless, all experimental values of the spin lifetimes are some orders of magnitude shorter than theoretically predicted~\cite{PhysRevLett.103.146801,
PhysRevB.80.041405} indicating that in present devices spin
transport is limited by extrinsic sources of spin scattering. These
include spin-orbit coupling by adatoms, edge effects and ripples.~\cite{PhysRevB.87.081402,PhysRevLett.103.146801,PhysRevB.87.075455,PhysRevLett.104.187201,PhysRevLett.103.026804,1367-2630-14-3-033015,Avsar2011}
Additionally, spin scattering may result from the underlying
substrate or the spin injection and detection contacts.~\cite{PhysRevB.80.041405,PhysRevLett.105.167202,PhysRevB.86.235408} The
importance of the latter might be indicated by recent electron spin
resonance (ESR) experiments on graphene nanoribbons and small flakes that were only weakly
coupled to the substrate and had no electrodes.~\cite{
doi:10.1021/nn302745x,AugustyniakJablokow2013118} Interestingly, the measured spin lifetimes
of localized spin states are at least \unit{200}{\nano\second} while the estimated spin lifetimes of conduction electrons are \unit{30}{\nano\second}, which is larger than any reported values from electrical Hanle spin precession measurements.

In this Rapid Communication, we investigate the influence of MgO
barriers on spin and charge transport properties by fabricating both
single layer (SLG) and bilayer graphene (BLG) non-local spin-valve
devices with variable contact resistance area products $R_{c}A$ of the MgO/Co
electrodes. We explore the relationship between spin lifetime $\tau_{\text{s}}$ and
charge carrier mobility $\mu$ in SLG and find a similar $1/\mu$ dependence
as seen in previous spin transport studies on exfoliated bilayer
graphene (BLG) devices.~\cite{PhysRevLett.107.047206} This
dependence is only seen in samples with $R_{c}A>\unit{1}{\kilo\ohm\micro\meter\squared}$. In fact, we observe that devices with long $\tau_s$ additionally exhibit a second Dirac peak in charge
transport, which stems from the electrodes. This
contact-induced Dirac peak overlaps with the Dirac peak of the bare
graphene which complicates the analysis of the carrier mobility and
thus complicates a clear assignment of the dominant spin scattering
mechanism in graphene. For devices with low $R_{c}A$ we
find an overall strong decrease of $\tau_s$ showing that
transparent contacts yield additional spin dephasing in graphene
underneath the contacts.

We have fabricated exfoliated SLG and BLG devices on
$\textrm{SiO}_2$(\unit{300}{\nano\meter})/$\textrm{Si}^{++}$ wafers.
The number of graphene layers is determined by optical contrast
measurement which is calibrated by Raman spectroscopy. After
e-beam lithography we use molecular beam epitaxy to first grow an MgO
spin injection/detection barrier with varying thicknesses from
\unit{1}{\nano\meter}  up to \unit{3}{\nano\meter} followed by
\unit{35}{\nano\meter} thick ferromagnetic Co contacts. The rather
thick barrier is necessary due to the fact that MgO on graphene
growths in the Volmer-Weber mode (island formation) if no wetting
layer is used.~\cite{wang:183107} We have evidence that even devices
with large $R_cA$ contacts that show non-linear differential
I-V-curves still exhibit pinholes in the barrier.~\footnote[1]{Paper in
preparation.} Thus the exact current distribution through the contact areas is unknown, which complicates the assignment of the correct $R_{c}A$ values.\cite{Nagashio2010} For the sake of simplicity, we assume a homogeneous current distribution for our analysis.

All transport measurements are performed under vacuum condition at
room temperature (RT) using standard lock-in techniques.~\footnote[2]{See
Supplemental Material for a more detailed description of device
fabrication, measurement methods and supporting measurements.} The
highly doped $\textrm{Si}^{++}$-wafer is used as a backgate which
allows changing the charge carrier density $n=\alpha
\left(V_{\textrm{G}} -V_{\textrm{D}}\right)$ in the graphene sheet
according to the established capacitor model~\cite{Novoselov2004}
with $\alpha \approx \unit{7.18 \cdot
\power{10}{10}}{\volt^{-1}\centi\meter\rpsquared}$, $V_{\textrm{G}}$
being the applied gate voltage and $V_{\textrm{D}}$ being the gate
voltage position of the maximum resistivity at the charge neutrality
point, also called Dirac point. By a linear fit of the conductance
$\sigma$, we extract the charge carrier mobility $\mu
=1/e\cdot\partial\sigma/\partial n$ at an electron density of
$n=\unit{1.5 \cdot \power{10}{12}}{\centi\meter\rpsquared}$. Hanle
spin precession measurements are performed in standard non-local
4-terminal geometry and are fitted by a simplified analytical
solution~\cite{ISI000249789600001,PhysRevB.37.5312} of the
steady-state Bloch-Torrey equation:~\footnotemark[2]
\begin{equation}
  \frac{\partial \vec{s}}{\partial t}\;=\;\vec{s}\times \vec{\omega}_0+D_{\text{s}}\nabla^2\vec{s}-\frac{\vec{s}}{\tau_{\text{s}}}\;=0,
\end{equation}
where $\vec{s}$ is the net spin vector, $\omega_0=g\mu_B B/\hbar$
the Larmor frequency, $\textit{D}_{\textrm{s}}$ the spin diffusion
constant and $\tau_{\text{s}}$ the transverse spin lifetime. Recent
experiments indicate that the effective g-factor in graphene-based
spin transport devices may differ from the free electron value at
low temperatures after a hydrogen
treatment.~\cite{PhysRevLett.109.186604} As ESR measurements for untreated graphene show $g\approx 2$ even for low temperatures~\cite{
doi:10.1021/nn302745x,AugustyniakJablokow2013118} and we also restrict ourselves to
RT, we assume $g=2$ for all devices in this study.

\begin{figure}[tbp]
\includegraphics{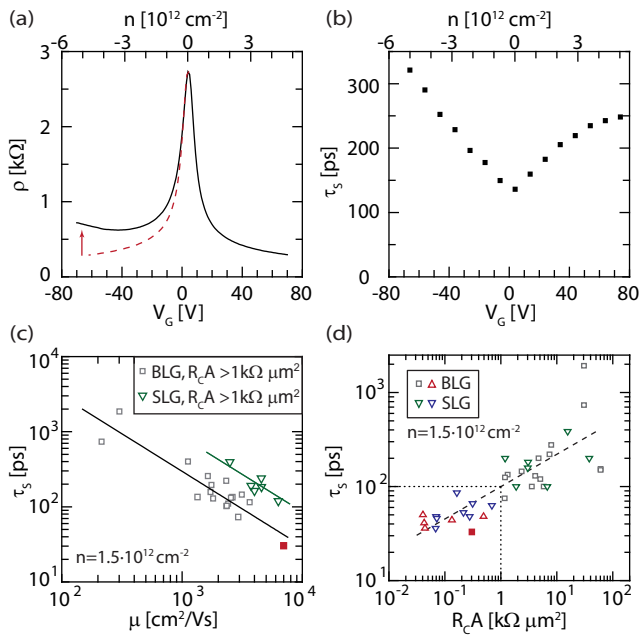}
\caption{(Color online) (a) $\rho$ vs $V_G$ of a SLG device with large
$R_{c}A$. The increase towards negative $V_G$
(see red arrow) indicates the existence of a second charge
neutrality point. As a guide to the eye the electron branch for $V_G
>0$ is mirrored at the Dirac point (red dashed line). (b)
$\tau_{\text{s}}$ vs $V_G$ and $n$; the minimum at the Dirac point is
typical for devices with large $R_{c}A$. (c) $\tau_{\text{s}}$ vs
$\mu$ at $n=\unit{1.5 \cdot \power{10}{12}}{\centi\meter\rpsquared}$
taken at RT for BLG (taken from
Ref.~\onlinecite{PhysRevLett.107.047206}) and SLG devices with large $R_{c}A$ contacts. The lines are the best fit to a DP like spin
dephasing. The BLG device with the shortest spin lifetime (filled square) will be discussed separately in the text. (d) $\tau_{\text{s}}$ vs $R_{c}A$ of all spin transport devices in (c) and Fig. 2(a).  }
    \label{fig:fig1}
\end{figure}

In Figs. 1(a)-(c) we show typical transport data for a SLG device
with $R_{c}A>\unit{1}{\kilo\ohm\micro\meter\squared}$.\footnotemark[2] We first note that there
is a strong electron-hole asymmetry in charge transport (Fig. 1(a))
as seen by the increase of the graphene resistivity for hole doping
towards large negative $V_G$ values. Its origin will be discussed
further below. Spin lifetimes are extracted from Hanle curves,\cite{PhysRevLett.107.047206}
which have been measured in perpendicular magnetic fields. The gate voltage
dependent $\tau_{\text{s}}$ times in Fig. 1(b) show a minimum at the Dirac
point and increase with both electron and hole doping. This
general trend is observed for most large $R_{c}A$ devices.

We next evaluate the dependence of $\tau_{\text{s}}$ on the electron
mobility $\mu$ at $n=\unit{1.5 \cdot
\power{10}{12}}{\centi\meter\rpsquared}$ for all SLG devices in
Fig.~1(c) (green triangles) on a log-log scale. For easier
comparison we include results on BLG (gray squares in Fig. 1(c)),
which some of us had previously
measured.~\cite{PhysRevLett.107.047206} The most striking
observation is that like in BLG $\tau_{\text{s}}$ depends inversely on
$\mu$ in our SLG devices. This relationship was previously
attributed to the dominance of D'yakonov-Perel' (DP) like spin
dephasing in graphene.

Remarkably, SLG devices exhibit longer spin lifetimes than BLG
devices of equal mobility. The vertical offset between SLG and BLG
in Fig. 1(c) can be analyzed within the DP spin dephasing mechanism.
For this we replace the momentum scattering time $\tau_{\textrm{m}}$
in the DP formula $1/\tau_{\textrm{s}} =
\Omega_{\textrm{eff}}^2\left( \Delta_{\textrm{SO}}\right)
\tau_{\textrm{m}}$ with the Boltzmann
expression of the mobility
$\mu=e\tau_{\textrm{m}}/m_{\textrm{eff}}^*$ and take the logarithm:~\cite{ISI000249789600001}
\begin{equation}
    \textrm{ln} \left( \tau_{\textrm{s}} \right) = \textrm{ln} \left( \frac{e}{\Omega_{\textrm{eff}}^2\left( \Delta_{\textrm{SO}}\right)\cdot{m_{\textrm{eff}}^*}} \right)  - \textrm{ln} \left( \mu\right),
\end{equation}
where \textit{e} is the elementary charge, $\Omega_{\textrm{eff}}^2$
the effective Larmor frequency which is dependent on the spin-orbit
coupling $\Delta_{\textrm{SO}}$ and $m_{\textrm{eff}}^*$ is the
effective mass. With this expression it is obvious that the vertical
offset in Fig.~1(c) can either result from a smaller effective mass
or a smaller overall spin-orbit coupling strength in the SLG
devices. We note that SLG is expected to exhibit massless
Dirac fermions near the Dirac point only in simple tight-binding
approximations. It has been shown that even the small intrinsic
spin-orbit coupling in SLG gives rise to a small effective
mass of the charge carriers,\cite{PhysRevB.80.235431,PhysRevB.82.245412} which supports our simple approach in Eq.~2. Even stronger
effects are expected from extrinsic sources such as contacts,
adatoms, and the underlying
substrate.~\cite{1367-2630-14-3-033015,PhysRevLett.103.026804,PhysRevB.80.041405}
As all experimental values of $\tau_{\text{s}}$ are well below theoretical
predictions, we expect that spin relaxation and dephasing is governed by extrinsic
sources in present devices. Because of the dominant extrinsic
contribution to the spin-orbit coupling slight changes in the
fabrication steps between the BLG and SLG devices (in our case another batch of
wafer, another resist for lithography) might be the reason for the
observed offset in the lifetime. These changes in sample fabrication may also explain the overall larger carrier mobilities in the new series of SLG devices seen in \ref{fig:fig1}(c) (no device under \unit{2000}{\centi\meter\squared\per\volt\second}).

In the following, we will focus on the influence of $R_{c}A$ on $\tau_s$. Several groups have suggested to use high resistive tunneling contacts to avoid the backflow of charge carrier spins into the ferromagnetic electrodes which otherwise yields a reduction of the spin lifetime.\cite{PhysRevB.80.214427,PhysRevLett.105.167202,Dlubak2012,PhysRevB.86.235408}
As mentioned above, all of our SLG devices exhibit large $R_{c}A$ values. Revisiting our previous BLG measurements also reveals $R_{c}A$ values above $\unit{1}{\kilo\ohm\micro\meter\squared}$ for almost all BLG data points in Fig. 1(c) (gray squares). Only the BLG device with
the highest mobility (red, filled square in Fig. 1(c)) has low $R_{c}A$ contacts with a flat differential
d$I$/d$V$-curve.~\footnotemark[2]

\begin{figure}[tbp]
\includegraphics{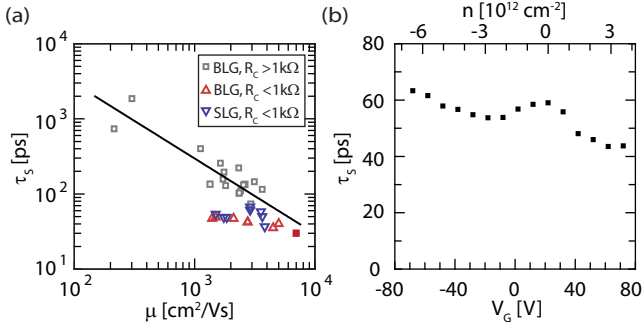}
\caption{(Color online) (a) $\tau_{\text{s}}$ vs $\mu$ taken at $n=\unit{1.5 \cdot \power{10}{12}}{\centi\meter\rpsquared}$ at \unit{300}{\kelvin}. Squares are taken from Ref.~\onlinecite{PhysRevLett.107.047206}; upward and downward pointing triangles are BLG and SLG devices, respectively. Both exhibit $R_{c}A<\unit{1}{\kilo\ohm\micro\meter\squared}$. (b) $\tau_{\text{s}}$ vs $V_G$ dependence of a typical device with transparent contacts.}
    \label{fig:fig2}
\end{figure}

At first sight this red data point seems to follow the DP like trend of the
large $R_{c}A$ BLG devices. To explore this in further detail we
fabricated additional SLG and BLG samples with a thinner MgO barrier but otherwise same fabrication
procedure. All of those show $R_{c}A<\unit{1}{\kilo\ohm\micro\meter\squared}$ (Fig.~1(d) blue and red data points). As seen in Fig.~2(a),
they exhibit strongly reduced $\tau_s$ values which vary between 30 and
\unit{70}{\pico\second} (upward pointing triangles for BLG and downward pointing ones for SLG) and lie well below all large $R_{c}A$ devices with no significant difference
between SLG and BLG devices. Furthermore, the data do not follow the
1/$\mu$ dependence. It is therefore obvious that devices with
$R_{c}A<\unit{1}{\kilo\ohm\micro\meter\squared}$ exhibit short spin lifetimes in which the above $\mu$ dependence of the large $R_{c}A$ devices is hidden by an additional spin dephasing channel which most likely results from the contacts.

The strong influence of low $R_{c}A$ contacts on the spin transport can
also be seen by the charge density dependence of $\tau_{\text{s}}$
(Fig.~2(b)), which is similar for all low $R_{c}A$ devices. In contrast
to all large $R_{c}A$ devices at room temperature (see Fig.~1(b)), $\tau_{\text{s}}$ does not
increase away from the Dirac point, but it rather decreases and may
increase again at larger carrier densities. Although we presently do
not understand this qualitative change in the density dependence, we
note that such a decrease of $\tau_{\text{s}}$ has previously also been observed in
BLG devices with large $R_{c}A$ contacts at low temperatures.~\cite{PhysRevLett.107.047206}

While we have seen that devices with low and large $R_{c}A$ values show a distinctly different mobility dependence of the spin lifetime (Fig.~1(c) and 2(a)), we now discuss the dependence of $\tau_s$ on $R_{c}A$ which is shown in Fig. 1(d) for all devices with measured $R_{c}A$ values. We note that even within a single device the respective $R_{c}A$ values for different contacts may vary significantly. The plotted $R_{c}A$ values in Fig.~1(d) are thus mean values of the respective injector and detector contacts of each device. Remarkably, we observe a significant increase of $\tau_s$ with $R_{c}A$ for all devices suggesting that the contacts are even the bottleneck for the large $R_{c}A$ devices. However, these devices also showed the pronounced inverse dependence of $\tau_s$ on $\mu$ (see Fig.~1(c)). It is therefore interesting to study if the influence of the contacts also becomes evident in charge transport.

\begin{figure}[tbp]
    \includegraphics{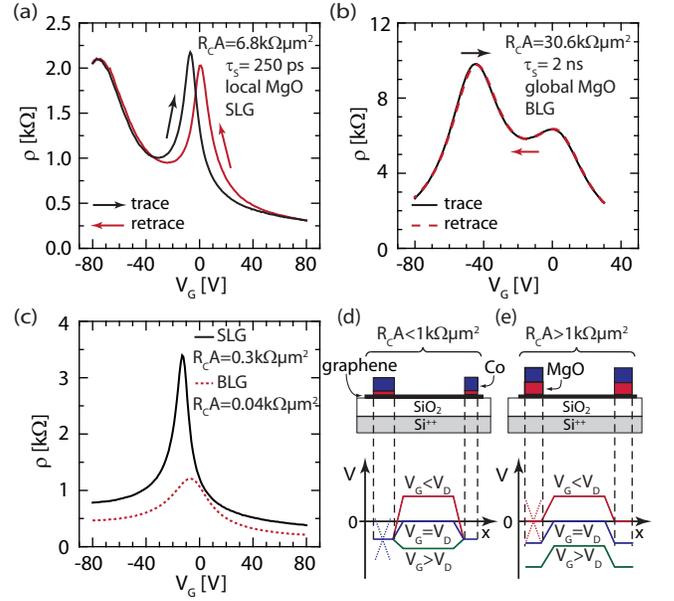}
\caption{(Color online) (a) $\rho$ vs $V_G$ of a large $R_{c}A$  SLG device with $\tau_{\text{s}}= \unit{250}{\pico\second}$ showing a
pronounced contact-induced second Dirac peak at $V_G=-75$~V. This
Dirac peak does not show any hysteresis in contrast to the Dirac
peak of the bare graphene near $V_G=0$~V. The arrows indicate the
sweep directions of the gate voltage. (b) $\rho$ vs $V_G$ for a large $R_{c}A$ BLG device with $\tau_{\text{s}}=2$~ns (see Fig.~2(a)) and
completely MgO covered graphene. No hysteresis is visible. (c)
$\rho$ vs $V_G$ of SLG and BLG devices with low $R_{c}A$ contacts.
Besides to a doping, the contacts may screen the gate field $V_G$
which can result in (d) pinning (low $R_{c}A$) or (e) no pinning (high
$R_{c}A$) of the electrostatic potential under the contacts
(corresponding Dirac cones indicated).}
    \label{fig:fig3}
\end{figure}

In Figs. 3(a) to (c), we summarize the gate voltage dependent graphene resistivity for both large and low $R_{c}A$ SLG and BLG devices. While we only observe one Dirac peak around zero gate voltage in all low $R_{c}A$ devices (see Fig.~3(c)),
we typically observe a second Dirac peak at larger negative gate
voltages for SLG and BLG devices with large $R_{c}A$ (Figs.~3(a) and (b)). Such a second Dirac peak has already been observed in spin-valve devices by another group.\cite{Vera-Marun2012}
The resistivity ratio of both Dirac peaks varies significantly from device to device. Not all
large $R_{c}A$ devices show the maximum of the second Dirac peak for
$V_{G}>\unit{-80}{\volt}$ (cp. to Fig.~1(a)). But the general trend is that the separation between both Dirac peaks gets smaller for devices with larger $R_{c}A$. Consistent with the additional scaling between $R_{c}A$ and $\tau_{\text{s}}$, the smallest peak separation has been observed in the device with the longest $\tau_{\text{s}}$ of \unit{2}{\nano\second} (Fig.~3(b)).

The left Dirac peak most likely results from the magnetic electrodes
while the right Dirac peak is due to charge transport
through the graphene sheet between the electrodes. This notion is
supported by hysteresis measurements when comparing different
devices with local and global MgO barriers. In the former case MgO
is only deposited underneath the ferromagnetic electrodes (see
Fig.~3(a) for corresponding SLG device) while in the latter case MgO
completely covers the graphene flake (see Fig.~3(b)). A hysteresis
is only observed for the right Dirac peak in the device with local
MgO barriers (Fig.~3(a)). It can originate from a thin water film on
top of the graphene flake.~\cite{Wang2010} Although we measure under
vacuum condition, such a hysteresis is initially always observed
before the water will eventually evaporate after a few hours.
However, no hysteresis for the left Dirac peak at negative gate
voltages is observed. If this peak results from the contact area,
this is also expected as water cannot cover the graphene underneath
the contact area. Consistent with this assignment, we do not observe
any hysteresis for global MgO devices (see Fig. 3(b)).

We next link the appearance of the second Dirac peak to the measured spin lifetime and the contact characteristics. It is well known that the contact material has a great influence on the transport properties in graphene. Scanning photocurrent microscopy experiments, for example, directly probe contact-induced doping and show Fermi level pinning from metallic electrodes.~\cite{PhysRevB.79.245430} A gate voltage dependent doping profile of the electrostatic potential $V$ for devices with low ohmic contacts is depicted in Fig. 3(d). Here $-e\cdot V$ equals to the position of the Fermi level in the graphene band structure. Although this profile can successfully explain an electron-hole-asymmetry in the resistivity,\cite{nouchi:253503} which we also observe in our low $R_{c}A$ devices, it cannot explain the second Dirac peak as the carrier density underneath the electrodes is not affected by the gate voltage.

As noted above, there is an island growth of our MgO barriers. In particular for thin barriers this favors the formation of conducting Co pinholes with presumably direct contact of the Co to the graphene layer. As shown by recent angle-resolved photoemission spectroscopy the Dirac cone of graphene is strongly shifted into the valence band for Co/graphene interfaces and its $\pi^*$ band hybridizes with 3d bands of Co near the Fermi level.\cite{Varykhalov2009,Varykhalov2012} This hybridization is consistent with the Fermi level pinning in our low $R_{c}A$ devices. It furthermore might account for the reduced spin lifetimes in Fig.~2(a) (red and blue data points) as the injected spins might be scattered by the 3d states in the graphene layer.

For large $R_{c}A$ devices with thicker MgO oxide barriers the formation of pinholes is suppressed.  Accordingly the 3d-hybridization of Co with graphene states gets diminished which can yield longer spin lifetimes. Furthermore, we expect a gradual depinning of the Fermi level. Together with the weaker Co induced n-doping with increasing oxide thickness\cite{PhysRevLett.101.026803,PhysRevB.87.075414} this also explains the appearance of the second Dirac peak. This situation is illustrated in Fig.~3(e) where the backgate voltage now also tunes the carrier density underneath the contacts. We note that the transition between pinning and depinning should in principal be continuous with increasing $R_{c}A$. In other words, the appearance of the second Dirac peak does not necessarily imply a complete depinning.
Considering the spatially inhomogeneous barrier thickness due to the Volmer-Weber island growth and remaining pinholes even for large $R_{c}A$ devices all current devices might not be in the regime of complete depinning. The rough Co/MgO interface may also result in inhomogeneous local magnetic fields which can be an additional source of spin dephasing.\cite{PhysRevB.84.054410} Finally, we do not observe a systematic dependence of the amplitude of the spin signal on the $R_{c}A$ values \footnotemark[2], which excludes a backflow of spins into the ferromagnetic electrode as a possible explanation of the observed $\tau_s$ dependence on $R_{c}A$.\cite{PhysRevB.80.214427,PhysRevLett.105.167202,PhysRevB.86.235408}

Next we address the calculation of $\mu$ for devices with a second Dirac peak. In Figs. 1(d) and 2(a) we determined $\mu$ and $n$ from the right Dirac peak which
we attribute to the bare graphene part. This might be a good
approach for devices with only one Dirac peak (Fig. 3(c)) or for
devices where the left Dirac peak is strongly separated in gate
voltage as in Fig.~1(a). In particular for devices with long $\tau_s$, however, the two Dirac peaks are not well separated but strongly overlap as seen in Figs.~3(a) and 3(b) for large $R_{c}A$
SLG and BLG devices, respectively. This overlapping has significant
influence on the slope $\partial\rho/\partial V_G$ of the right
Dirac peak at $n=\unit{1.5 \cdot
\power{10}{12}}{\centi\meter\rpsquared}$
($V_G-V_D\approx\unit{20}{\volt}$). The smaller the separation
between both Dirac peaks becomes the smaller the respective slope
and carrier mobility will be. We note that this results in an
underestimation of the mobility of the bare graphene part. It is
important to emphasize that the contact-induced left Dirac peak
might thus partially be responsible for the decrease in observed
carrier mobility for devices with longer spin lifetimes.

There are more elaborated models to determine carrier mobilities
including contact-induced pinning and depinning of the Fermi level
and the respective potential profiles of the graphene along the
device.~\cite{thiele:094505,6054021,JJAP.50.070109,
PhysRevLett.107.156601, Vera-Marun2012} However, there are too many
unknown quantities which currently hinder to extract reliable values
for the respective carrier mobilities in the different graphene
parts from a single gate dependent resistivity measurement.\footnotemark[2] Without
further measurements of the potential profile we thus cannot give a
more precise evaluation of the influence of the contact induced
Dirac peak on the carrier mobility. This, on the other hand, would
be important for identifying intrinsic spin dephasing mechanisms in
graphene. Our findings show that the understanding of spin transport
in graphene based non-local spin-valve devices requires independent
understanding of both spin and charge transport properties which may
significantly differ in graphene underneath the spin injection and
detection electrodes and graphene between the
electrodes.~\cite{doi:10.1021/nl301050a}

In summary, we have studied spin and charge transport in
graphene-based non-local spin-valves by tuning the $R_{c}A$ values of MgO
injection/detection barriers. For low $R_{c}A$ contacts, there is a
significant spin dephasing in graphene underneath the contacts,
while SLG and BLG devices with large $R_{c}A$ values show long
spin lifetimes at RT. The latter devices exhibit
a second Dirac peak at negative gate voltages. As the peak
separation is smallest for devices with the longest spin lifetimes,
it might partially account for the observed $1/\mu$
dependence of $\tau_{\text{s}}$.

This work was supported by DFG through FOR 912.

\selectlanguage{english}

\end{document}